\documentstyle[12pt]{article}

\begin{document}


\pagestyle{empty}

\begin{center}

\null\vspace{0.2in}

{\Large\bf GRAPE-6: A Petaflops Prototype}

\vspace{0.6in}

{\large\em Piet Hut}\\
\vspace{0.1in}
{\em School of Natural Sciences}\\
{\em Institute for Advanced Study, Princeton, NJ 08540}

\vspace{0.3in}

{\large\em Jeffrey M. Arnold}\\
\vspace{0.1in}
{\em 10686 Mira Lago Ter.}
{\em San Diego, CA 92131}

\vspace{0.3in}

{\large\em Junichiro Makino}\\
\vspace{0.1in}
{\em Department of General Systems Study, College of Arts and Sciences}\\
{\em University of Tokyo, 3-8-1 Komaba, Meguro-ku, Tokyo 153, Japan}

\vspace{0.3in}

{\large\em Stephen L. W. McMillan}\\
\vspace{0.1in}
{\em Department of Physics and Atmospheric Science}\\
{\em Drexel University, Philadelphia, PA 19104}

\vspace{0.3in}

{\large\em Thomas L. Sterling}\\
\vspace{0.1in}
{\em High Performance Computing Systems Group, Jet Propulsion Laboratory}\\
{\em Mail Stop 168-522, 4800 Oak Grove Dr., Pasadena, CA 91109}\\
\vspace{0.1in}
{\em Center for Advanced Computing Research, California Institute of
Technology}\\
{\em Mail Code 158-79, 1200 E. California Blvd., Pasadena, CA 91125}

\vspace{0.7in}

For email addresses and further information, see:
\vspace{0.01in}
http://casc.physics.drexel.edu

\vspace{1.0in}

Invited paper

1997 Petaflops Algorithms Workshop

April 13-18, 1997

Williamsburg, Virginia

\vfill

\end{center}


\newpage

\null\vspace{0.1in}

\begin{center}
\large\bf Abstract
\end{center}

\vspace{0.2in}

\noindent
We present the outline of a research project aimed at designing and
constructing a hybrid computing system that can be easily scaled up to
petaflops speeds.  As a first step, we envision building a prototype
which will consist 
of three main components: a general-purpose, programmable front end, a
special-purpose, fully hardwired computing engine, and a
multi-purpose, reconfigurable system.  The driving application will be
a suite of particle-based large-scale simulations in various areas of
physics.  The prototype system will achieve performance in the $\sim50$--$100$
teraflops range for a broad class of applications in this area.

\vspace{0.2in}

Central to our design is the isomorphic relation between the abstract
and the physical dataflow paths: the pipelined architecture of
the hardware will mimic the physical interactions in the simulation.
This tight binding between algorithm and architecture is the key to
obtaining maximum efficiency and throughput, with negligible latency and
hardly any overhead.

\vspace{0.2in}

The merging of custom LSI and reconfigurable logic will result in a
unique capability in performance and generality, combining the
extremely high throughput of special-purpose devices (SPD) with the
flexibility of reconfigurable structures.  The prototype system will
represent proof of concept of this new hybrid architecture.
A central goal is to explore performance
scalability, by dynamically adjusting the balance of workload and
interconnection overhead between the reconfigurable SPD and its host
computing system.  Issues of direct concern are the scaling with
problem size of parallelism, critical path, intra-system communication
bandwidth, and overhead.  This research will investigate the scaling
sensitivities of the role of the host computer both as system size
increases and as SPD computational demands vary due to the changes in
the SPD structure.

\vspace{0.2in}

We have been led to our overall design by the following
considerations.  The combination of a hardwired petaflops-class
computational engine and a front end with sustained speed on the order
of 10 gigaflops can produce extremely high performance, but only for
the limited class of problems in which there exists a single
bottleneck with computing cost dominating the total.  While the
calculation for which the Grape-4 (our system's immediate predecessor)
was designed is a prime example of such
a problem, in many other applications the primary computational
bottleneck, while still related to an inverse-square (gravitational,
Coulomb, etc.)  force, requires less than 99\% of the computing power.
Although the remainder of the CPU time is typically dominated by just
one secondary bottleneck, its nature varies greatly from problem to
problem.  It is not cost-effective to attempt to design custom chips
for each new problem that arises.  
FPGA-based systems can restore the balance, guaranteeing scalability
 from the teraflops to the petaflops domain, while still retaining
significant flexibility.

\vfill


\newpage
\setcounter{page}{1}
\pagestyle{plain}

\section*{1.~~Special-Purpose Computers in Scientific
		 Computation}

Many problems in computational science have the characteristic that
the bulk of the CPU time is spent in one or a few relatively short
sections of code.  One general area in which these problems are well
known and particularly acute is particle simulations.  Some examples
are:

\begin{itemize}

\item astrophysical simulations of stellar systems, in which particles
interact predominantly by means of long-range gravitational forces,

\item molecular dynamical (MD) simulations, which may involve both
long-range (Coulomb) and short-range (van der Waals) forces (e.g.~the
protein-folding problem; MD simulations in Materials Science),

\item kernel methods in fluid dynamics (e.g.~Smooth Particle
Hydrodynamics---SPH), in which local thermodynamic quantities such as
density, pressure, and specific energy are determined by sampling the
properties of nearby particles,

\item plasma-physics applications (ranging from the dynamics of the
solar wind to controlled fusion and the simulation of explosive
processes in astrophysics and elsewhere),

\item vortex methods in fluid simulations, in which individual vortex
elements are modeled as particles with well-defined interparticle
forces.

\end{itemize}

If these important applications are to achieve very high
(petaflops-class) performance, it is essential to find ways to
overcome the computational bottlenecks that presently limit their
utility.  We have set out to explore the possibilities of combining
general-purpose computers, custom-designed processors, and
reconfigurable logic elements to achieve this goal.  Although our
description is cast largely in terms of particle simulations, the
general physical context motivating this study, the results of our
work will have applicability to a far broader range of problems.

In all of the examples listed above, typical programs spend most of
their time computing a very small kernel, such as the determination of
particle interactions, that contains relatively few instructions but
processes a large amount of data.  This kernel is often sufficiently
simple that one can contemplate realizing it in the form of custom
hardware: a specially designed board, reconfigurable chip, or custom
LSI device.  The efficiencies inherent in hardware implementations of
algorithms, as well as the opportunities for massive parallelism, mean
that special-purpose systems can outperform general-purpose machines
by several orders of magnitude within the ranges of their
applicability, and can be designed and built on a time scale far
shorter than is normally feasible for commercial supercomputers.

In some contexts, one can achieve spectacular performance improvements
by concentrating the design effort on accelerating (with custom
hardware) the principal bottleneck in the problem.  However, it is
more usually the case that the cost of the next most expensive part of
the calculation is significant---a few percent of the total, say---so
the speedup one can achieve with the combination of special-purpose
hardware and a general-purpose front end is still rather limited.  For
example, in MD calculations of organic molecules, over 90\% (but less
than 99\%) of the calculation cost is spent on Coulomb forces, while
more than 90\% of the remainder of the calculation cost is spent in
computing van der Waals and other short-range forces.  Again, in
simulations of stars or galaxies with gas dynamics included, more than
90\% of the CPU time is typically spent on the gravity calculation,
with over 90\% of the rest spent on gas dynamics.

Rather than attempting to construct separate custom hardware devices
to handle all eventualities, our strategy will be to define a hybrid
system in which the principle bottleneck---in our case, the
computation of long-range, inverse-square forces---is accelerated with
custom hardware, but where the various secondary bottlenecks are
addressed using reconfigurable Field Programmable Gate Array (FPGA)
technology.  The design of such a
system combines custom high-performance
chip design, reconfigurable logic systems design, system architecture,
and algorithm development.  In the next section we describe the two
previous projects that form the basis for this collaboration: the
teraflops class special-purpose Grape-4, and the reconfigurable Splash
2.  We then proceed to describe in more detail the approach to be
followed and some specifics of the architecture.

\section*{2.~~Technical Background}

In 1995 and 1996, a series of workshops explored the extreme regime of
trans-teraflops computing to near-petaflops-scale performance. These
community-led forums involved experts from academia, industry, the
national labs, and government in a diversity of fields including
device technology, computer architecture, system software, and
application algorithms. The broad objective of this initiative was to
identify the opportunities and challenges towards achieving petaflops
computing and to formulate a research program that will lead this
nation aggressively toward that goal.

It was found that using conventional COTS technologies anticipated
 from industry would require approximately 20 years to deliver
petaflops scale execution rates, but that alternative methods could
deliver such capability in less than half that time. An important
conclusion was that special-purpose devices, possibly augmented with
reconfigurable logic structures, were capable of being the first to
reach this goal and could achieve petaflops scale performance for
mission-critical applications within as little as 5 years, and
at moderate cost. It was recommended that reconfigurable special
purpose devices be pursued for well defined and highly important
applications and that methods for doing so be explored. Our research
program directly addresses this recommendation and focuses on the
domain-specific problem of particle simulations.

\subsection*{2.1~~Particle Dynamics and the Grape Project}

This project is the natural advancement of a class of
special-purpose devices whose heritage extends back to the Digital
Orrery designed and built by G. Sussman (of MIT, then visiting
Caltech) and coworkers (Applegate et al.~1985) to perform very
high-precision, long-term integration of particle problems.  This
original work focused on the evolution and stability of the solar
system, but the techniques extend to a wide array of particle-based
applications.  This early research motivated an international
collaboration which developed the Grape series of special-purpose
computers that ultimately achieved teraflops-class sustained
performance, winning Gordon Bell awards in both 1995 and 1996.

The purpose of the Grape project is the long-term integration of large
self-gravitating systems, such as star clusters and galaxies---the
celebrated ``gravitational $N$-body problem'' in its fullest form.
While significant speed-up has been obtained in certain circumstances
through the introduction of efficient algorithms (e.g.~Barnes and Hut
1986, 1989; Greengard \& Rohklin 1987), this problem continues to tax
the capabilities of even the fastest general-purpose machines.
Detailed analysis of the best algorithms available for the study of
dense stellar systems (Hut, Makino \& McMillan 1988; Makino \& Hut
1988, 1989) indicated that speeds on the order 1 Tflops would be
required to model even a small star cluster with any degree of
realism.

In 1989, a team of researchers (led by D.~Sugimoto, of Tokyo
University) began to explore the feasibility of building
special-purpose hardware for stellar dynamics simulations.  Familiar
with the earlier success of the Orrery, and aware of the disappointing
performance of stellar dynamics calculations on general-purpose
machines, they realized that a novel approach might be appropriate.
To this end, they constructed a series of special-purpose
``accelerators'' to speed up critical parts of the simulations.  The
first machine in the series, known as Grape-1, was completed in the
Fall of 1989 (Ito et al.~1990).  (The acronym ``Grape'' stands for
GRAvity PipE, and designates a very efficient hardware implementation
of the Newtonian pairwise force between particles in a
self-gravitating $N$-body system.)  Grape-1 had a speed of over 200
Mflops, but relatively low ($\sim1$\%) precision.  Just six years
later, the Grape-4 achieved a speed of over 1 Tflops, with high
(15 decimal digit) precision (Makino et al.~1997).

Constructed specifically for astrophysical simulations, the Grape-4
consists of 1692 custom LSI chips, each of which computes the
gravitational interaction between two particles (in this case, stars).
The design of the Grape hardware is such that, once all pipelines are
filled, each chip produces one new interparticle interaction
(corresponding to approximately 60 floating-point operations) every
three clock cycles.  With a clock speed of 30 MHz, a peak chip speed of
0.6 Gflops is achieved.  Operating all chips in parallel gives a
theoretical peak speed of 1.08 Tflops.

 From a programming standpoint, the Grape simply replaces a section of
existing code by a series of hardware calls, implemented as library
functions, that return the desired information, namely, the force on a
specified particle or group of particles.  The distribution of
computing load is such that the ~1 Tflops Grape-4 can be driven at
better than 50\% of peak speed by a 100--200 Mflops host (which
performs the rest of the dynamical simulation), for systems comprising
20,000--50,000 particles.

\subsection*{2.2~~Splash 2}

The second critical component of this project is the use of
reconfigurable logic to enable high-performance, but flexible,
implementations of application-driven algorithms.  The Splash project
(Buell et al.~1996, Arnold et al.~1992, Arnold 1995) was designed to
provide exceptional performance on a general range of systolic
problems.  The first generation Splash system consisted of a fixed
length linear systolic array of Xilinx 3090 FPGAs.  Although
successful for many algorithms, the limited I/O bandwidth, the
inflexible interprocessor communication and the difficulty of
programming Splash 1 prevented wider use.

The Splash 2 system connected sixteen Xilinx 4010 FPGAs in a linear
array augmented with a crossbar data path to increase the flexibility
of the communication network.  The system could be scaled up to 256
FPGAs by extending the linear array across multiple boards.  Each FPGA
in the array was coupled to an independent memory element for local
storage.  Splash 2 pioneered the use of a simulation based programming
methodology which created a very rich application development
environment.  

A wide variety of applications were written for the Splash 2 system,
including database searches, macromolecule sequence analysis, and real
time signal and image processing.  It was discovered that the most
successful of these applications exhibited a small number of common
themes.  These themes included: streams of small data objects which
required relatively low precision arithmetic, such as pixel streams
for image processing; very long pipelines spanning many FPGAs, such as
the macromolecule sequence comparison applications; static
communication patterns among the processing elements that did not
require dynamic routing overhead, such as the FFT algorithm.

Splash 2 successfully demonstrated the utility of reconfigurable
computing to reduce the overhead associated with fixed instruction set
computers.  Direct hardware implementation of algorithms in
reconfigurable logic eliminates the overhead of instruction stream
interpretation.  Reconfigurable logic allows the programmer to tailor
the width of data path elements to match the precision requirements of
the application, resulting in significant area and performance
savings.  Partial evaluation techniques can be used to fold constant
values into the logic, further reducing the silicon real estate
required by an application.

\section*{3.~~Integration of SPD and FPGA Systems}

\subsection*{3.1~~Algorithms}

Particle-based simulations are eminently suitable for a wide class of
physical problems.  Unlike grid-based methods, particle methods are
well suited to the treatment to extremely inhomogeneous systems with
large time-dependent density contrasts.  Also, there is no problem
with mesh tangling, regridding or subgridding: particles sampling the
fluid (either in physical space or in phase space) form a naturally
comoving system of reference points.

Within the general class of particle simulations, there is a rich
variety of integration algorithms, and a large set of problem-specific
interparticle forces.  Designing custom LSI chips for each separate
problem and algorithm is not feasible.  A much better alternative is
to use reconfigurable hardware to establish domain-specific pipeline
structures for whole classes of problems, programming the detailed
algorithmic implementation directly into the FPGA subsystem.  The best
of all worlds, however, is to combine the two approaches.

Even though most particle-based simulations have their own specific
types of interparticle forces, in many cases there is at least one
force component that scales as the inverse square of the distance.
Gravity is an obvious example of a fundamental force with this
property; electrostatics is another.  Vortex methods in
hydrodynamics offer an example of a derived type of force field, in
which vortex elements can also be approximated as point particles
which obey inverse-square interactions.

The common functionality of interparticle force laws across much of
the class of particle simulations supports the development of custom
LSI chips to work in tandem with FPGA-based systems on which the
remainder of the force interactions can be performed.  In fact, since
the inverse-square component of the force dominates the total
computational cost in many cases, the use of custom LSI chips does
more than simplifying the overall design: it speeds up the total
throughput considerably.

\subsection*{3.2~~Implementation}

Hardware acceleration of critical segments of a computation allows the
high cost/performance of special-purpose hardware to be combined with
the flexibility of existing workstations without the need for special
software development.  This approach can be compared to using
hand-coded assembly-language or machine-code for an inner loop in an
algorithm that itself is programmed in a higher-level language---the
difference being that this inner loop is now realized directly in
silicon.

Whether or not it is feasible to implement a complex interparticle
force law, such as SPH, using FPGA technology depends critically on
the relative accuracy required.  Preliminary studies suggest that the
required relative accuracy is actually rather low in many cases, and
an error of a few percent may be acceptable. In this case, it is
possible to implement a full SPH pipeline in a single 1997 FPGA chip;
multiple chips may be used to implement more complex pipelines.  It
should soon be possible to implement several pipelines in a single
larger chip, although the increase in density of FPGA devices is not
as rapid as that of gate arrays.  Some commercially available systems
may already be close to satisfying this need.

Reconfigurable computing engines based on FPGA technology appear very
well suited for implementation of SPH and other interactions.  A
natural question is: why not proceed one step further, and implement
the entire system, including the functionality of the custom hardware,
using FPGA?  The answer is that the performance that can be achieved
with FPGA is at least two orders of magnitude lower than that can be
achieved by custom LSIs. There is about a factor of 100 difference in
the available circuit densities between FPGAs and custom LSIs.  In
addition, there is a factor of few difference in the clock
frequencies.  These factors have been roughly constant since the first
FPGA was introduced, and are likely to remain roughly constant
in the future.  Thus, if we were to use FPGAs as the basic building
blocks for our system, our cost/performance would be
degraded by a factor of 100, placing the cost of a full petaflops
system close to \$1 billion.

\section*{4.~~Scalability and Systems Integration}

System scalability is dependent on a number of closely associated
architectural attributes. One is the ratio of the intrinsic
algorithmic parallelism to the critical path length as the problem
size grows. A second is the overhead required in managing the larger
problem size which itself may contribute to the critical path
length. A third scalability issue for systems of the type described here is
the balance of workload between the special-purpose device and the
host as problem size increases. This last issue, if not properly
handled, can defeat the purpose of employing special-purpose
subsystems as Amdahl's law takes effect.

As the total system workload increases due to problem size, the amount
of work to be performed by the host, the frequency at which this work
needs to be done, and the amount of information that must be exchanged
between the host and the special-purpose device will all contribute to
establishing the upper bound to scalability. In addition, the amount
of overlap between processing by the special-purpose device and the host
can extend the apparent scalability by trading parallelism for
latency, thus hiding the execution time of the host subsystem. All
these factors vary with problem size.

The basic means of addressing the scalability issues associated
with the system-level architecture are:

\begin{itemize}

\item increase the speed and capacity of the host,

\item increase the bandwidth between host and SPD,\footnote{In the
general discussion of this subsection
only, we use the term SPD to refer to the entire system,
including the FPGA part.}

\item reduce latency times of interconnect between host and SPD

\item increase the proportion of total workload performed by the SPD, and

\item organize the work profile so that both host and SPD are
      operating concurrently.

\end{itemize}

\noindent
The size of the host can be expanded through
conventional parallel MIMD organization. Note that this requires that the
workload of the host exhibit
sufficient coarse-grain parallelism to exploit the resources of the
MIMD architecture.

The intra-system bandwidth and latency poses an increasing
challenge. While specific optical technologies for communication are
or will be in the near future capable of Gbps throughput, the
bottlenecks may very well be the interface to the host system
itself. Algorithms must be redefined to minimize the information
flow between the two subsystems by
transferring an increasing proportion of the
total workload onto the SPD. This has proved to be necessary with more
conventional accelerators used to augment conventional system
capabilities in such areas as signal and image processing. 
By configuring the SPD to provide adequate buffering and flow control to
enable concurrent operation of both systems, it is possible to
effectively take the host system out of
the critical path and permit the SPD to operate at peak
performance.

An important contribution of this project is the
inclusion of reconfigurable logic within the structure of the
special-purpose device. The objective is to broaden the class and
generality of algorithms that may be performed by the system, thus
increasing its utility and cost effectiveness. The use of
reconfigurable logic within the SPD can significantly alter
the computational and timing relationships between the host and the
SPD. As the internal topology of the
SPD is modified through reconfiguration, the SPD workload profile
will change. This is
particularly true where use of data paths may be conditional (dependent
on intermediate values) causing the execution profile to vary.

Without associated changes to the host processor workload, execution
time could be dominated by the host and scalability constrained, again
by Amdahl's law. However, since the new structures enabled by the
reconfigurable logic imply new applications and algorithms, the programmer
will be redefining the host codes as well. A negative aspect of this
is that the careful balancing of host to SPD computation is not done
just once, as in the case of conventional fixed SPDs, but many times
as the adaptive SPD is imbued with new application algorithms.

We are exploring effective ways of
mitigating the challenges of repeatedly having to program the host and
balance its computational demands with that of the reconfigurable
SPD. Software tools must be adapted or developed to assist in
this special case of heterogeneous computing to expose the demands and
behaviors of the new algorithms as they relate to the system workload
balance.

\section*{5.~~Architecture}

Our planned architecture is a high-performance hybrid computing
system consisting of three components: a general purpose front end
processor; a multi-purpose reconfigurable back end processor array;
and a custom designed, special-purpose back end processor array.  The
system will provide a direct means of investigating the
behavior of reconfigurable systems at performance levels, and on
problem sizes, sufficient to advance the state of knowledge in an
important scientific discipline.

\subsection*{5.1~~Custom Chip Performance}

The present Grape-4 processor represents 1990/1991 technology (1$\mu$m
fabrication line width).  Even if no changes were made in the basic
design, we estimate that advances in fabrication technology would
permit more transistors per chip and increased clock speed, enabling a
50-100 MHz, 10-30 Gflops chip in a 1996 start (0.35 $\mu$m line
width), and a 100--200 MHz, 50--200 Gflops chip with 1998 (0.25 $\mu$m)
technology.  Based on these projected performance improvements, we
conclude that $\sim$10,000 chips of $\sim$100 Gflops each could be
combined to achieve petaflops speeds by the year 2000.  (This is one
of several conclusions reached by a recently completed NSF Point
Design study in which we were involved.)
Fabrication of
250--500 copies of the chip will enable system design at
the board (16 chips per board) and controller (16--32 boards per
controller) levels.  Anticipated performance is
in the 50--100 Tflops range.  The design of the force-calculation
pipeline in the new processor chip is essentially similar to that in
the Grape-4, although a number of changes will be made to optimize the
design further.  A $\sim$10-Gflops front end will allow the
system to run at close to peak speed.

\subsection*{5.2~~System Structure}

Our architecture can be viewed as a collection of hardware
function accelerators attached to a general purpose front end computer
through a high speed network.  In this system there are two categories
of accelerator, or ``back end'', processor: a fixed function processor
for computing inverse-square law forces; and a reconfigurable
processor capable of computing a wide variety of functions.  The fixed
function processors will be implemented as the custom LSI chips just
described.  The reconfigurable processors will be based upon
commercially available FPGA devices, with an anticipated sustained
performance of 1--10 Gflops, depending upon the application.

The overall organization of the system has yet to be determined, but
one possibility is:

\begin{itemize}

\item Multiple pipelines per processor.  For the custom processors,
sixteen inverse-square pipelines per chip should be possible.  For the
reconfigurable processors this number will be
application dependent.

\item Sixteen processors per board.

\item Up to 32 boards per controller.  This level of the hierarchy is
called a {\em cluster}.

\item Initially, two to four clusters per front-end host.  This number
is scalable to 16 clusters, corresponding to a 1.2 Pflops machine.

\end{itemize}

A high-performance point-to-point network will
implement the cluster-level interconnect.  This network will likely
extend down to the level of the individual processors and up to the
host.

To achieve the processor-to-memory bandwidth required to sustain the
inverse-square law processors it will be necessary to integrate the
particle data memory into the processor chip.  Using 30\% of the
total silicon as SRAM, one chip can have at least 2 Mbits of SRAM
memory. One particle requires about 600 bits of storage, so
one chip can store at least $3\times 10^3$ particles. Committing 1/3
of the silicon for memory, we need around 12,000 chips to achieve 1
Pflops. Thus, the total number of particles which can be stored is
$4\times 10^7$, which is more than the maximum number of particles for
which direct-summation algorithms are practical.

With the FPGA technology expected to be available in
1998, and the software techniques described earlier, it should be
possible to build reconfigurable pipelines capable of delivering in
excess of 1 Gflops performance.  Much more problematic will be the
delivery of data to those pipelines to sustain that level of
performance.  The particle force pipelines envisioned typically
consume on the order of 20 bytes of data per clock cycle.  Assuming a
clock rate of 100~MHz, keeping the pipeline filled will require an
FPGA-to-memory bandwidth of the order of 2.5~GB/sec.  To approach this
level of performance will require a very wide path to a
synchronous DRAM.

Most of the force calculation pipelines currently envisioned are
expected to fit comfortably into an FPGA of the density expected to be
available in 1998.  However, some applications may
require deeper pipelines than will fit in a single FPGA.  The
decision of whether to interconnect the FPGA processors to permit a
single pipeline to span multiple FPGAs is difficult.  The obvious
advantage would be to increase the range of applicability of the
reconfigurable processors.  There are several significant
disadvantages, however, including using I/O pins that could otherwise
be devoted to memory bandwidth, and inserting I/O pad drive and receive
times to the critical path of the application pipeline.

For a typical application the system will operate as follows:

\begin{itemize}

\item The front end loads the FPGA configuration corresponding to the
function to be computed into the reconfigurable back end processors.

\item The front end loads the set of particle descriptors into the
back end memories.

\item Each back end processor reads the particle data, computes the
forces on each of a list of target particles, and writes the modified
particle data back to memory.

\item The front end retrieves the results, computes the next time
step, and iterates.

\end{itemize}

\subsection*{5.3~~Outstanding Questions}

Several important questions must be addressed:

\begin{itemize}

\item FPGA architecture.  There are several competing device
architectures available commercially with advantages and disadvantages
to each.  A detailed analysis of options is required.

\item Memory organization.  To achieve the required bandwidth it will
be necessary to look to high performance memory interfaces such as
Rambus or synchronous DRAM.

\item Network topology.  The design of the communication network is
another critical aspect of the hardware architecture.  Although a
hierarchy of bus-based connections was used in Grape-4, in our
system some form of point-to-point structure will be required
to achieve the necessary bandwidth.  Tree and sorting network
topologies are under study.

\item Interprocessor interconnect.  Ideally, an integral number of
application pipelines would fit in a single FPGA, in which case the
only interprocessor communication occurs through the particle memory.
However, some applications may not fit in a single FPGA and require
spanning across multiple chips.

\item Ratio of fixed to reconfigurable computing resources.  The
optimal mix of these two paradigms depends on both the computational
mix of the application and on the relative costs.

\end{itemize}

\noindent
Factors to consider in the architecture study include: 1)
the level of technology available commercially; 2) the balance of
memory size to logic density; 3) the required memory bandwidth per
unit of logic for the expected application set; 4) the expected
size and nature of the application pipelines; and 5) the performance
of the FPGA I/O.

\section*{6.~~Software Issues}

\subsection*{6.1.~~Software Tools and Methodologies}

One of the key advantages of reconfigurable computing is the ability
to tailor the hardware to match the requirements of the application.
In the case of the numerical applications in which we are most interested,
reconfigurable logic will allow the application programmer to control
the data path widths to exactly match the arithmetic range and
precision requirements of the algorithm.  This will result in a
substantial savings of real estate and improved performance for those
portions of the algorithm which do not require high precision.
Additional savings can be achieved by partial evaluation of arithmetic
expressions to fold constant values into the synthesized logic.

Partial evaluation and the exploitation of varying precision and range
in an application will require the development of sophisticated
analysis tools.  One tool will derive lower bounds on the data path
width of every arithmetic operator in a function from a set of input
and output constraints.  It will work in conjunction with a
set of parametrized module generators capable of building efficient
arithmetic operators over a wide range of size and throughput
requirements.

The methodology for mapping algorithms onto the reconfigurable
processors will rely on commercial CAD technology wherever possible.
Most applications will use logic synthesis from a high-level design
language such as VHDL, although for efficiency some
commonly used critical structures such as adders and
multipliers will be designed  at a lower level.  Final mapping to the
particular FPGA
technology will be provided by the device vendor's tool suite.  

Development of the reconfigurable application
modules will be facilitated by a simulation environment consisting
of a common interface specification for the
FPGA processor and behavioral models of the surrounding circuitry.
This will allow the application code to be simulated in the context of
the surrounding system while still providing all of the features of a
source level debugger.

A number of software challenges exist for the front-end host computer.
The device driver and runtime library must support very high bandwidth
data movement between the user data space and the back end memory.
Efficient synchronization primitives must be built to coordinate the
interaction between the user code and the back end.  A debugger must
be developed to support the development of the reconfigurable code.
This debugger should allow single stepping and internal register state
examination of the FPGAs.

\subsection*{6.2
~~Software Mapping Strategy}

A critical aspect of our architecture is the
programmability of the reconfigurable back-end processors. The
functions to be mapped onto reconfigurable logic will be expressed
as a set of arithmetic expressions augmented with annotations on the
arguments and results which specify the range and precision
requirements.  This textual description is input to an arithmetic
analysis tool
which produces a synthesizable VHDL model of the function.  The
functionality of the resulting model may be verified through
simulation before synthesis and technology mapping into FPGA
bitstreams.

The arithmetic analysis tool converts an annotated mathematical
expression into a synthesizable VHDL circuit.  The input to the tool
specifies the data format of the arguments and the range and precision
of the results.  The constraints represented by the annotations is
propagated through the data flow graph of the expression to determine
the minimum range and precision required of each of the operators.
The resulting set of labeled operators is then used to construct an
application specific module library.  For each unique combination of
operator and parameter set, a module generator for that operator type
is invoked with the appropriate parameters.  The module generator
produces a synthesizable VHDL model of the specific operator, which is
added to the library.  A top level structural VHDL description of a
circuit which implements the data flow graph in terms of calls to
elements in the module library is then generated.  Finally, the
resulting VHDL model is synthesized by conventional logic synthesis
tools to form a gate list suitable for input to the FPGA vendor's
mapping tools.

A key component of this mapping strategy is a library of parametrized
fixed and floating point module generators.  These generators are
parametrized on the precision and range of their arguments and
results, allowing the creation of data path elements and
representation formats that exactly match the requirements of the
algorithm.  The modules produced by the generators
obey a common specification, allowing the composition of modules
into long pipelines.

\subsection*{6.3~~Runtime Software}

We have ready at hand a suite of well-tested codes for large-scale
particle-based simulations.  Some of these have been developed by
others, and subsequently modified and extended by us, while other
codes have been developed from scratch.  Our most sophisticated code,
named Kira, has been designed and developed over the last few years
(with the Grape-4 as a development platform)
with the explicit purpose of treating simultaneously problems that
span huge ranges of time scales and length scales.

The original motivation stemmed from astrophysics.  In order to follow
the complete evolution of a star cluster, from soon after the Big Bang
to the present, we have to model a history that spans across ten
billion years.  During this period, however, we have to resolve
occasional critical episodes, during which highly energetic processes
in double stars can take place on time scales of milliseconds (when
neutron stars or black holes are involved).  Our algorithm thus has to
deal with time scales that span a range of $10^{20}$.  Similarly,
physical length scales range from a hundred light years down to
kilometers, spanning a range of $10^{13}$.

None of the orbit-integration algorithms found in standard texts on
the solution of ordinary differential equations can handle these
extreme requirements.  Our solution has been to implement individual
time step sizes that are continually adjusted for each particle.
Tightly interacting subgroups of particles are treated locally by
constructing dynamically changing recursively refined coordinate
patches.  Most importantly, on all levels hooks are included that
allow us to transparently interface with independent software modules
that treat additional physical processes.  Some of these
non-gravitational effects, such as stellar evolution, will be run on
the host, while the more compute-intensive interactions, such as those
involving hydrodynamics, will be run on the FPGA part of our system.

In addition to these astrophysically motivated problems, the Kira code
can be modified to treat other problems in physics that require
modeling through particle realizations.  The modularity and
flexibility of the Kira architecture will make it straightforward to
change the force-law module, and to modify the particle-interaction
management separately.

\section*{7.~~Comparison with Other Work}

Only a small number of large-scale reconfigurable systems have been
built to date.  The most notable of these efforts are the DEC Paris
Research Lab's PeRLe projects; the IDA SRC's Splash project (described
elsewhere in this proposal; Buell et al.~1996, Arnold et al.~1992,
Arnold 1995); and HP Lab's ongoing Teramac project.

The PeRLe systems (Bertin \& Touati 1994, Vuillemin et al.~1996) were
designed to act as a tightly coupled general purpose configurable
hardware coprocessor.  The PeRLe programming model consisted of a
large array of bit level functional units called ``programmable active
memory'' cells implemented in an array of Xilinx 3000 FPGAs.  A number
of applications were written for PeRLe with impressive results: a 50
tap 16-bit FIR filter ran at 16 times real time audio rate.  An RSA
decryption implementation outperformed the then state-of-the-art
custom ASIC by an order of magnitude.  PeRLe and its successors have
been used for a number of high energy physics applications at CERN.

Amerson et al. (1995, 1996) at Hewlett-Packard Labs have built a large
scale reconfigurable computing engine called Teramac, based upon a
custom FPGA design called Plasma.  Teramac, which can execute
synchronous logic designs of up to one million gates, was designed to
conduct experiments in using special purpose processors to search
large non-text databases and to perform volume visualization.

Because of the traditionally low logic density of FPGAs, there have
been relatively few efforts to map floating point intensive
applications onto reconfigurable hardware.  Shirazi et al. (1995)
implemented a 512-point Fast Fourier Transform for image processing on
Splash 2 using an 18-bit floating point format.  The 10 bit mantissa
and 7 bit exponent provided sufficient precision and range for the
image stream of 8-bit pixels.  The resulting pipelined multiplier
occupied 44\% of the Xilinx XC4010-6 (Xilinx 1994) FPGA (1991 technology)
while the adder occupied about 28\%.  One tap of a FIR filter composed
of these units fit in a single XC4010 and delivered 20Mflops.

Cook et al.~(1995) have implemented an N-body accelerator on the
Altera RIPP-10 board which contains eight Altera FLEX81188 FPGAs plus
2MB of static RAM.  This accelerator employs a direct method
inverse-square law force calculator pipelined across all eight FPGAs.
The arithmetic implemented (Louca et al.~1996) is single precision IEEE
compatible floating point.  Using a digit-serial approach, Louca was
able to achieve a peak performance of 2.5 Mflops per floating point
multiplier, which is the rate limiting operator.  The entire system
of 8 FPGAs runs at 10 MHz to deliver 2.5 million force calculations per
second, for a peak performance of 40 Mflops.

In the area of special-purpose machines for computational science,
only QCD systems such as QCDSP and APE offer performance comparable to
the Grape series.  QCDSP, a 4th generation special-purpose QCD machine
designed by Christ and coworkers at Columbia University, should be
completed soon.  It will consist of 16384 TI DSP chips (50 Mflops
each, for 800 Gflops theoretical peak and 200--300 Gflops actual
performance), with a total cost of some \$3 million.  The current
generation of the Italian APE project (APE-100) should ultimately
consist of 2048 custom chips each delivering 50 Mflops peak
performance.  Of these, 1024 have been installed so far.  The
next-generation system (APEmille) will have a total budget of \$13
million and performance in the 1--2 Tflops range.

The Grape-4 and the proposed new LSI chip are superior in
performance to these current and planned QCD systems.  On the other
hand, the QCD machines offer some degree of programmability, and
therefore have broader applicability than Grape-4.  However, the
hybrid architecture in Grape-6 significantly widens the range
of application of our system.

General-purpose systems are in many senses complementary to our
project, as we need a high-speed, general-purpose
computer as a front end.  However, the performance requirements for
our host system are in the tens of gigaflops, rather than the
multi-teraflops range that will represent the state of the art in
2000.  No general-purpose system currently under consideration (and,
specifically, none of the general-purpose systems funded by the
present NSF Petaflops Point Design program) is likely to achieve
speeds in the 100 Tflops--1~Pflops range before around 2010.  The
probable leaders in general-purpose high-performance computation over
the next few years, the massively parallel ASCI Red and Blue systems,
will have peak speeds in the 2--3 Tflops range, with longer-range
targets of 10 Tflops by 1999 and 100 Tflops by 2002.

\section*{8.~~Summary}

The project described here significantly advances the state of
the art in the application of reconfigurable logic to computational
science by developing new methods and architectural structures through
the direct design and implementation of a proof-of-concept system. The
driving premise is that hybrid structures of
custom high-density processing components and programmable
reconfigurable logic devices can be used to realize physical
structures with the cost and performance of special-purpose systems
and the flexibility of significantly more general systems.

The bases for this significant advance are
the Splash 2 and Grape-4 projects, which
represent leading developments in reconfigurable and
special-purpose systems, respectively.  We expect significant
advances in system architecture, algorithm to hardware layout,
software tools, and hardware structures to come from this work. The
resultant techniques and devices
will directly impact achievable
computation of a specific class of problems, and will provide the means
of realizing other comparable system structures for a wide range of
computational domains.

The application driven focus of our research is the general class of
particle simulation problems that range from astrophysics and
cosmology to molecular dynamics and may relate to plasmas,
magneto-hydrodynamics, materials, computational fluid dynamics, and
many other realms of scientific and engineering inquiry. Of these,
gravitational simulation, molecular dynamics, and CFD will be pursued
as detailed areas of concentration to apply and evaluate the concepts
and technologies to be developed as part of this research project.


\section*{Bibliography}

\noindent
Amerson, R. et al. 1995, in Proc. IEEE Symposium on FPGAs for Custom
Computing Machines (Los Alamitos, CA: IEEE Comp. Soc. Press), p. 32.

\noindent
Amerson, R. et al. 1996, in ACM/SIGDA Int. Symp. on Field Programmable
Gate Arrays, Monterey, CA.

\noindent
J. H. Applegate, M. R. Douglas, Y. G\"usel, P.  Hunter, C. L. Seitz,
and G. J. Sussman, {\it IEEE Transactions on Computers}, Vol C34, p
822, 1985.

\noindent
Arnold, J.M, Buell, D.A. \& Davis, E.G. 1992, in ACM Symp. Parallel
Algorithms and Architectures (New York: ACM Press), p. 316.

\noindent
Arnold, J.M. 1995, J. of Supercomputing, 9, 277.

\noindent
Barnes, J. \& Hut, P. 1986, Nature, 324, 446

\noindent
Barnes, J. \& Hut, P. 1989, Astroph. J. Supp., 70, 389

\noindent
Bertin, P. \& Touati, H. 1994, in Proc. IEEE Symposium on FPGAs for
Custom Computing Machines (Los Alamitos, CA: IEEE Comp. Soc. Press), p. 133.

\noindent
Buell, D.A., Arnold, J.M., Kleinfelder, W.J. 1996, Splash 2: FPGAs in a Custom
Computing Machine (Los Alamitos, CA: IEEE Comp. Soc. Press).

\noindent
Cook, T.A., Kim, H-R. \& Louca, L. 1995, in SPIE Photonics East
Conference on Field Programmable Gate Arrays for Fast Board
Development and Reconfigurable Computing, p 115.

\noindent
Greengard, L. \& Rohklin, V. 1987, J. Comp. Phys. 73, 325.

\noindent
Hut, P., Makino, J. \& McMillan, S. L. W. 1988, Nature 336, 31.

\noindent
Ito T., Makino, J., Ebisuzaki, T.  \& Sugimoto, D. 1990, 
Comp. Phys. Comm., 60, 187.

\noindent
Louca, L., Cook, T.A. \& Johnson, W.H. 1996, in
Proc. IEEE Symposium on FPGAs for Custom Computing Machines,
(Los Alamitos, CA: IEEE Comp. Soc. Press), p. 107.

\noindent
Makino, J. \& Hut, P.  1988, Astroph. J. Supp., 68, 833.

\noindent
Makino, J. \& Hut, P., 1989, Comp. Phys. Rep. 9, 199.

\noindent
Makino, J., Taiji, M., Ebisuzaki, T., \& Sugimoto, D. 1997,
Astroph. J., 480, xxx.

\noindent
Shirazi, N., Walters, A. \& Athanas, P. 1995, in
Proc. IEEE Symposium on FPGAs for Custom Computing Machines
(Los Alamitos, CA: IEEE Comp. Soc. Press), p. 153.

\noindent
Vuillemin et al., J. 1996, IEEE Trans. VLSI Systems, March 1996.

\noindent
Xilinx, Inc., 1994, The Programmable Gate Array Data Book, San Jose, CA.

\end{document}